\documentclass[aps,prb,twocolumn,superscriptaddress,showpacs,floatfix]{revtex4}

\usepackage{graphicx}
\usepackage{dcolumn}
\usepackage{bm}
\usepackage{stmaryrd}
\usepackage{latexsym}
\usepackage{amssymb}
\usepackage{amsfonts}
\usepackage{amsmath}

\begin{document}
\title{Collective excitations in one-dimensional ultracold Fermi gases: a comparative study}
\date{\today}

\author{Wei Li}
\affiliation{Department of Physics, Zhejiang Normal University, Jinhua, Zhejiang Province, 321004, China}
\author{Gao Xianlong}
\email{gaoxl@zjnu.cn}
\affiliation{Department of Physics, Zhejiang Normal University, Jinhua, Zhejiang Province, 321004, China}
\author{Corinna Kollath}
\affiliation{Centre de Physique Th\'eorique, Ecole Polytechnique,CNRS, 91128
 Palaiseau Cedex, France}
\author{Marco Polini}
\affiliation{NEST-CNR-INFM and Scuola Normale Superiore, I-56126 Pisa, Italy}
\affiliation{Department of Physics, Zhejiang Normal University, Jinhua, Zhejiang Province, 321004, China}

\begin{abstract}
Time-dependent density-functional theory (TDDFT) is a powerful 
tool to study the non-equilibrium dynamics of inhomogeneous interacting many-body systems. 
Here we show that the simple adiabatic local-spin-density approximation for the time-dependent exchange-correlation potential is surprisingly accurate in describing collective density and spin dynamics in strongly correlated one-dimensional ultracold Fermi gases. Our conclusions are based on extensive comparisons between our TDDFT results and 
accurate results based on the adaptive time-dependent density-matrix renormalization-group method.
\end{abstract}
\pacs{71.15.Mb, 71.10.Pm}
\maketitle

\maketitle

\section{Introduction}

Quantum many-body systems of one-dimensional (1D) interacting particles have attracted an enormous interest for more than fifty years~\cite{giamarchi_book}. These systems are nowadays available in a large number of different laboratory realizations ranging from single-wall carbon nanotubes~\cite{Saito_book} to semiconductor nanowires~\cite{yacoby_sc_separation}, conducting molecules~\cite{Nitzan_2003}, chiral Luttinger liquids at fractional quantum Hall edges~\cite{xLL}, and trapped atomic gases~\cite{rmp_cold_atoms,moritz_2005,greiner_2001}.  

Regardless of statistics, the effective low-energy description of many of these systems is based on a harmonic theory of long-wavelength fluctuations~\cite{haldane}, {\it i.e.} on the ``Luttinger liquid" model~\cite{giamarchi_book}. 
One distinctive feature of the Luttinger liquid is that its low-energy spectrum is completely dominated by 
collective excitations as opposed to individual quasiparticles that carry both charge and spin. A single-particle excitation in a Luttinger liquid directly decays into collective spin and charge excitations that propagate with different velocities. This phenomenon is called ``spin-charge separation"~\cite{giamarchi_book}.

Tunneling measurements between two parallel quantum wires in GaAs/AlGaAs heterostructures with varying electron density have demonstrated~\cite{yacoby_sc_separation} the existence of collective spin and charge excitations with different velocities. 

It has also been proposed to study experimentally the dynamics of spin and charge excitations {\it in real time} 
using 1D two-component cold Fermi gases~\cite{moritz_2005},  where ``spin" and ``charge" refer, respectively, to two internal (hyperfine) atomic states and to the atomic mass density~\cite{recati_prl_2003,kecke_prl_2005,kollath_prl_2005,kollath_jpb_2006, kollath_njp_2006}. In Ref.~\onlinecite{polini_PRL_2007} a different aspect of this collective behavior has been pointed out: namely, spin excitations are intrinsically damped at finite temperature, while charge excitations are not.

Important for the experimental observation of the time-evolution of excitations in ultracold quantum gases is the creation of a sizable perturbation of the gas (strong enough to be detected). The theoretical description of such strong perturbations, however, needs techniques going beyond the low-energy Luttinger-liquid model~\cite{giamarchi_book}. 
For example, the decay of sizable density perturbations~\cite{kollath_prl_2005} and of a single-particle excitation~\cite{kollath_jpb_2006, kollath_njp_2006} has been recently demonstrated in real time in a numerical time-dependent density-matrix renormalization-group (tDMRG)~\cite{feiguin,daley} study of the 1D Fermi-Hubbard model.

Aim of the present work is to provide a conceptually and numerically simple time-dependent microscopic 
many-body theory that is capable of capturing the main physical features of the time-evolution of collective excitations. 

A powerful theoretical tool to study the interplay between interactions and time-dependent 
inhomogeneous external fields of arbitrary shape is the time-dependent density-functional theory (TDDFT)~\cite{Giuliani_and_Vignale,vignale_kohn, marques_2006}, which is based on the Runge-Gross theorem~\cite{rgt} and on the time-dependent Kohn-Sham equations. Many-body effects enter TDDFT {\it via} the time-dependent exchange-correlation (xc) functional, which is often treated by the {\it adiabatic} local-density approximation (ALDA)~\cite{Giuliani_and_Vignale, zangwill}. In this approximation one assumes that the time-dependent xc potential is just the {\it static} xc potential evaluated at the instantaneous density. The static xc potential is then treated within the static LDA. The main characteristic of the ALDA is that it is local in time, as well as in space. Memory effects, whereby the xc potential at a time instant might depend on the density at an earlier time, are completely ignored. Very attractive features of the ALDA are its extreme simplicity, the ease of implementation, and the fact that it is {\it not} restricted to small deviations from the ground-state density, {\it i.e.} to the linear-response regime. 

Even though several non-adiabatic beyond-ALDA approximate functionals are available nowadays (see {\it e.g.} Refs.~\onlinecite{vk_1996,dobson_1997,vuc_1997, tokatly_2005,orestes_2007}), in this work we focus on a simple adiabatic xc functional and test its performance in describing a particular problem: collective density and spin dynamics in strongly correlated inhomogeneous lattice systems. Building upon earlier ideas described at length in Refs.~\onlinecite{soft,capelle,gao_prb_2006,schenk_condmat_2008,verdozzi_2007}, we here employ a lattice TDDFT scheme in which the time-dependent xc potential is determined {\it exactly} at the adiabatic local-spin-density-approximation level through the Bethe-{\it Ansatz} solution of the homogeneous 1D Hubbard model~\cite{lieb_wu}. The numerical results based on this scheme are tested against accurate adaptive tDMRG simulation data for both spin-unpolarized and spin-polarized systems.

The contents of the paper are briefly described as follows. In Sect.~\ref{sect:model} we introduce the model 
lattice Hamiltonian that we use to study collective excitations and we briefly summarize its exact solution in the absence of external potentials. In Sect.~\ref{sect:tSOFT} we present the self-consistent lattice TDDFT scheme that we use to deal with the time-dependent inhomogeneous system and introduce the Bethe-{\it Ansatz} adiabatic local-spin-density approximation that we employ for the xc potential. In Sect.~\ref{sect:numerical_results} we report and discuss our main numerical results in comparison with tDMRG simulation data. Finally, in Sect.~\ref{sect:conclusions} we summarize our main conclusions and future perspectives. 

\section{The model}
\label{sect:model}

We consider a two-component repulsive Fermi gas with 
$N$ atoms confined to a 1D tube and subjected to an optical lattice 
potential applied in the direction of the tube. The optical lattice has unit lattice constant and $L$ 
lattice sites. For times $t \leq 0$ the system is in the presence of a spin-selective 
focused laser-induced potential, which creates a strong local disturbance in the ultracold gas. 
At time $t=0^+$ this local potential is {\it suddenly} turned off: we are interested in the subsequent time evolution of the spin-resolved densities. 

This system is modeled by the following Fermi-Hubbard Hamiltonian~\cite{jaksch_1998},
\begin{eqnarray}\label{eq:hubbard}
	{\hat {\cal H}}(t)&=&-\gamma\sum_{i,\sigma}({\hat c}^{\dagger}_{i\sigma}
	{\hat c}_{i+1\sigma}+{\rm H}.{\rm c}.)+U\sum_i 
	{\hat n}_{i\uparrow}{\hat n}_{i\downarrow}\nonumber\\
	&+&\sum_{i, \sigma} V_{i\sigma}(t){\hat n}_{i\sigma} \equiv {\hat {\cal H}}_{\rm ref}+{\hat {\cal H}}_{\rm ext}(t)~.
\end{eqnarray}
In Eq.~(\ref{eq:hubbard}), $\gamma$ is the hopping parameter, ${\hat c}^{\dagger}_{i\sigma}$ (${\hat c}_{i\sigma}$) creates (destroys)
a fermion on the $i$th site ($i \in[1,L]$), 
$\sigma=\uparrow,\downarrow$ is a pseudospin-$1/2$ degree-of-freedom 
(hyperfine-state label), $U>0$ is the strength of the on-site Hubbard repulsion, and
${\hat n}_{i\sigma}={\hat c}^{\dagger}_{i\sigma}{\hat c}_{i\sigma}$. We also introduce for future purposes the local number operator ${\hat n}_{i}={\hat n}_{i\uparrow}+{\hat n}_{i\downarrow}$ and the local spin operator ${\hat s}_{i}={\hat n}_{i\uparrow}-{\hat n}_{i\downarrow}$. 

The ``time-dependent" Hamiltonian ${\hat {\cal H}}_{\rm ext}(t)$ models the aforementioned spin-selective focused laser-induced potential. 
The external potential $V_{i\sigma}(t)$ is taken to be of the following simple Gaussian form
\begin{eqnarray}\label{eq:ext_pot}
V_{i\sigma}(t)&=&W_\sigma \exp{\left\{-\frac{[i-(L+1)/2]^2}{2w^2}\right\}} \Theta(-t)\nonumber\\
&\equiv& V^{\rm ext}_{i\sigma} \Theta(-t)~,
\end{eqnarray}
where $\Theta(x)$ is the Heaviside step function. This 
guarantees that the local potential $V^{\rm ext}_{i\sigma}$, which is active for all times $t\leq 0$, 
is suddenly switched-off at time $t=0^+$. Note that the time $t$ enters the problem through the step function only. 
We are not really studying a time-dependent problem but only the dynamics of a system after a sudden local quench: an initial state $|\Psi_0\rangle$, 
which is an eigenstate of ${\hat {\cal H}}_{\rm ref}+\sum_{i, \sigma} V^{\rm ext}_{i\sigma} {\hat n}_{i\sigma}$, is propagated forward in time with a different Hamiltonian 
(${\hat {\cal H}}_{\rm ref}$). $|\Psi(t)\rangle= \exp{(-i {\hat {\cal H}}_{\rm ref} t)}|\Psi_0\rangle$ 
is the state of the system at time $t$. Present-day technology in cold-atom-gas laboratories 
allows to change the external potentials 
on short time scales. By this it is possible to explore the regime where the many-body system is still 
governed by a unitary evolution but with non-equilibrium initial conditions. 

The number of atoms with spin up, $N_\uparrow=\sum_i \langle \Psi(t)|{\hat n}_{i\uparrow} |\Psi(t) \rangle$, 
can be different from the number of atoms with spin down,
$N_\downarrow=\sum_i \langle \Psi(t)| {\hat n}_{i\downarrow} |\Psi(t)\rangle$. The particle number $N_\uparrow$ and $N_\downarrow$ are separately conserved quantities because 
no spin-flip mechanism is included in the Hamiltonian (\ref{eq:hubbard}). The model (\ref{eq:hubbard}) must be accompanied by some boundary conditions: in this work we choose for simplicity open (hard wall) boundary conditions (OBCs). 
OBCs do not model well the most common experimental set-ups~\cite{rmp_cold_atoms,moritz_2005,greiner_2001} in which a parabolic trapping acts on the Fermi gas to keep it into the optical lattice. However, we have deliberately decided to limit our present investigations to simple OBCs to disentangle spurious effects (mainly spatial coexistence of different quantum phases) that can be induced by the parabolic trapping from fundamental effects related to the dynamics after the quench.

In the absence of ${\hat {\cal H}}_{\rm ext}(t)$ [{\it i.e.} for $V_{i\sigma}(t)=0$], 
the Hamiltonian in Eq.~(\ref{eq:hubbard}) reduces to a 1D homogeneous 
Hubbard model  that has been solved exactly by Lieb and Wu~\cite{lieb_wu}. 
At zero temperature the properties of ${\hat {\cal H}}_{\rm ref}$ in the thermodynamic limit ($N_\sigma, L \rightarrow \infty$) are determined by the spin-resolved filling factors $n_\sigma=N_\sigma/L$ and by the dimensionless coupling constant $u \equiv U/\gamma$. 
For simplicity, we limit the analysis below to $n=n_\uparrow+n_\downarrow<1$ and, for definiteness, we take $n_\uparrow \ge n_\downarrow$.

According to Lieb and Wu~\cite{lieb_wu}, 
the ground state of ${\hat {\cal H}}_{\rm ref}$ in the presence of repulsive interactions and 
in the thermodynamic limit is described by two continuous distribution functions $\rho(x)$ and $\sigma(y)$ which satisfy the Bethe-{\it Ansatz} (BA) coupled integral equations,
\begin{eqnarray}\label{eq:lw_1}
\rho(x)=\frac{1}{2\pi}+\frac{\cos{x}}{\pi}\int_{-B}^{+B}\frac{u/4}{(u/4)^2+(y-\sin{x})^2}\sigma(y)dy
\end{eqnarray}
and
\begin{eqnarray}\label{eq:lw_2}
\sigma(y)&=&\frac{1}{\pi}\int_{-Q}^{+Q}\frac{u/4}{(u/4)^2+(y-\sin{x})^2}\rho(x)dx\nonumber\\
&-&\frac{1}{\pi}
\int_{-B}^{+B}\frac{u/2}{(u/2)^2+(y-y')^2}\sigma(y')dy'\,.
\end{eqnarray}
The parameters $Q$ and $B$ are determined by the normalization conditions
\begin{equation}\label{eq:lw_3}
\left\{
\begin{array}{l}
{\displaystyle \int_{-Q}^{+Q}\rho(x)dx=n}\vspace{0.1 cm}\\
{\displaystyle \int_{-B}^{+B}\sigma(y)dy=n_\downarrow}
\end{array}
\right.~.
\end{equation} 
The ground-state energy of the system (per site) is given by
\begin{equation}\label{eq:gsenergy}
\varepsilon_{\rm GS}(n_\uparrow, n_\downarrow, u)=-2 \gamma \int_{-Q}^{+Q}\rho(x)\cos{x}~dx~.
\end{equation}

\section{Time-evolution within time-dependent density-functional theory}
\label{sect:tSOFT}

In this Section we describe the two-step procedure that we have followed to calculate the time-evolution of the spin-resolved 
site-occupation profiles $n_{i\sigma}(t)$. We first calculate the spin-resolved 
site-occupation profiles corresponding to $|\Psi_0\rangle$ 
for times $t\leq 0$ by means of a static DFT calculation and then find 
the subsequent time evolution for $t>0$ by means of TDDFT.

\subsection{Preparation of the initial state}
\label{sect:gs}

As we have already noted, for times $t \leq 0$ the Hamiltonian (\ref{eq:hubbard}) with OBCs 
describes an equilibrium ground-state problem. 
We can calculate accurately the ground-state spin-resolved site-occupation profiles of the inhomogeneous system 
described by ${\hat {\cal H}}(t\leq 0)={\hat {\cal H}}_{\rm ref}+\sum_{i, \sigma} V^{\rm ext}_{i\sigma}
{\hat n}_{i\sigma}$ by means of a DFT scheme based 
on Eqs.~(\ref{eq:lw_1})-(\ref{eq:gsenergy}). We have in fact generalized the site-occupation-functional theory (SOFT) scheme proposed in Ref.~\onlinecite{gao_prb_2006} and based on the BA local-density approximation 
to the case in which the external potential is spin-dependent (spin-SOFT). 
We here summarize the main steps that we have followed 
to calculate the ground-state spin-resolved site-occupation profiles $n_{i\sigma}(t\leq 0)$. 

Within spin-SOFT $n_{i\sigma} \equiv n_{i\sigma}(t\leq 0)$
can be obtained by solving self-consistently the static lattice Kohn-Sham (KS) equations
\begin{equation}\label{eq:sks}
\sum_{j}[-\gamma_{ij}+V^{\rm KS}_{i\sigma}\delta_{ij}]\varphi^{(\alpha)}_{j\sigma}=\varepsilon^{(\alpha)}_\sigma \varphi^{(\alpha)}_{i\sigma}
\end{equation}
 together with the closure
\begin{equation}\label{eq:closure}
n_{i\sigma}=\sum_{\alpha, {\rm occ.}}\left|\varphi^{(\alpha)}_{i\sigma}\right|^2~,
\end{equation}
where the sum runs over the occupied orbitals. Here 
$\gamma_{ij} = \gamma > 0$ if $i$ and $j$ are nearest-neighbor sites and zero otherwise and 
$V^{\rm KS}_{i\sigma}=U n_{i{\bar \sigma}}+V^{\rm xc}_{i\sigma}+V^{\rm ext}_{i\sigma}$ where ${\bar \sigma}=-\sigma$. The first term in the effective Kohn-Sham potential $V^{\rm KS}_{i\sigma}$ is the Hartree mean-field contribution, while $V^{\rm xc}_{i\sigma}$ is the (not exactly known) xc potential. As already stressed in Ref.~\onlinecite{gao_prb_2006}, exchange interactions between parallel-pseudospin atoms have been effectively eliminated in the Hubbard model~(\ref{eq:hubbard}) by restricting the model to one orbital per site. Hence parallel-pseudospin interactions are not treated dynamically in solving the Hamiltonian, but are accounted for implicitly {\it via} a restriction of the Hilbert space. To stress the analogy of the present work with {\it ab initio} applications of standard DFT, we nevertheless continue to call $V^{\rm xc}_{i\sigma}$ the exchange-correlation potential, but it is understood that the exchange contribution to this quantity is exactly zero.

The local-spin-density approximation (LSDA) has been shown to provide an excellent account of the ground-state properties of a large variety of inhomogeneous systems~\cite{d&g,joulbert_1998,Giuliani_and_Vignale}. In this work we have employed 
the following BA-based LSDA (BA-LSDA) functional
\begin{equation}\label{eq:balsda}
V^{\rm xc}_{i\sigma}\simeq \left.V^{\rm xc}_{i\sigma}\right|_{\rm BA-LSDA} \equiv 
\left. v^{\rm hom}_{{\rm xc}, \sigma}(n_\uparrow,n_\downarrow,u)\right|_{n_\sigma \rightarrow n_{i\sigma}}\,,
\end{equation}
where, in analogy with {\it ab initio} spin-DFT, the xc potential $v^{\rm hom}_{{\rm xc}, \sigma}(n_\uparrow,n_\downarrow,u)$ 
of the reference system described by ${\hat {\cal H}}_{\rm ref}$ is defined by
\begin{eqnarray}\label{eq:vxchom}
v^{\rm hom}_{{\rm xc}, \sigma}(n_\uparrow,n_\downarrow,u)&=&
\frac{\partial}{\partial n_\sigma}\left[\varepsilon_{\rm GS}(n_\uparrow,n_\downarrow,u)\right.\nonumber\\
&-&\left.\varepsilon_{\rm GS}(n_\uparrow,n_\downarrow,0)-Un_\sigma n_{\bar \sigma}\right]~.
\end{eqnarray}
Thus, within the LSDA scheme proposed in Eqs.~(\ref{eq:balsda}) and~(\ref{eq:vxchom}), the only necessary input is the xc potential $v^{\rm hom}_{{\rm xc}, \sigma}(n_\uparrow,n_\downarrow,u)$ of the homogeneous reference system, which 
can be calculated exactly following a very similar procedure to that outlined in Ref.~\onlinecite{gao_prb_2006}.

The self-consistent scheme represented by Eqs.~(\ref{eq:sks})-(\ref{eq:vxchom}) can be solved numerically for each set of 
parameters $\{L,N_\uparrow,N_\downarrow,u,W_\sigma/\gamma,w\}$. 
The outcome of these calculations is  $n_{i\sigma}$, which is used in the next section as the initial condition for the time evolution. 

\subsection{Time-evolution within TDDFT}
\label{sect:time-evol}

The spin-resolved site-occupation profiles
of the many-body system described by the Hamiltonian (\ref{eq:hubbard}) at time $t$ 
can be obtained by solving single-particle-like time-dependent lattice KS equations
\begin{equation}\label{eq:td-sks}
i\hbar \partial_t \psi^{(\alpha)}_{i\sigma}(t)=
\sum_{j}\left[-\gamma_{i,j}+V^{\rm KS}_{i\sigma}(t)\delta_{ij}\right]\psi^{(\alpha)}_{j\sigma}(t)
\end{equation}
with initial conditions $n_{i\sigma}(0)=n_{i\sigma}$. As in the static case, the spin-resolved time-dependent site occupations $n_{i\sigma}(t)$ are calculated by adding up the contributions of the orbitals that are occupied at the initial time,
\begin{equation}\label{eq:closure_dyn}
n_{i\sigma}(t)=\sum_{\alpha, {\rm occ.}}\left|\psi^{(\alpha)}_{i\sigma}(t)\right|^2~.
\end{equation}
In Eq.~(\ref{eq:td-sks}) $V^{\rm KS}_{i\sigma}(t)=U n_{i{\bar \sigma}}(t)+V^{\rm xc}_{i\sigma}(t)$ is the spin-resolved 
KS. 
The first term in the effective KS potential is the instantaneous Hartree mean-field contribution, while $V^{\rm xc}_{i\sigma}(t)$ is the (not exactly known) xc potential. The time-dependent spin-resolved KS potential must be determined self-consistently with the site-occupation profiles $n_{i\sigma}(t)$. This means, in practice, that the initial ground-state densities $n_{i\sigma}$ determine the initial KS potential, 
which is then used to recalculate the site-occupation profiles at an infinitesimally later time, and so on. 

In this work we have chosen to approximate the time-dependent xc potential $V^{\rm xc}_{i\sigma}(t)$  
with a BA-based adiabatic local-spin-density approximation (BA-ALSDA):
\begin{eqnarray}\label{eq:balsda_dyn}
V^{\rm xc}_{i\sigma}(t) &\simeq& \left.V^{\rm xc}_{i\sigma}(t)\right|_{\rm BA-ALSDA} \nonumber\\
&\equiv& 
\left. v^{\rm hom}_{{\rm xc}, \sigma}(n_\uparrow,n_\downarrow,u)\right|_{n_\sigma \rightarrow n_{i\sigma}(t)}~.
\end{eqnarray}

\section{Numerical results and discussion}
\label{sect:numerical_results}

In this Section we report some illustrative numerical results that we have obtained applying the TDDFT/BA-ALSDA 
method described above. 

All the numerical results presented below correspond to a system with $N=28$ atoms on 
$L=72$ sites, the OBCs being imposed at the sites $i=0$ and $i=73$.
We compare the results obtained with the TDDFT/BA-ALSDA with results of the adaptive tDMRG. The adaptive tDMRG relies on the use of an effective Hilbert space of dimension $M$ adapted at each time-step. In the current work the time-evolution is performed using a Suzuki-Trotter decomposition. Several hundred DMRG states are kept (up to $M=800$) and time-steps of the order of ${\cal O}(0.1~\hbar/\gamma)$ are used to achieve accurate results up to long times. For a discussion of the sources of uncertainties we refer the reader to the work by Gobert {\it et al.}~\cite{Gobert}.  

\begin{figure*}
\begin{center}
\tabcolsep=0cm
\begin{tabular}{cc}
\includegraphics[width=0.50\linewidth]{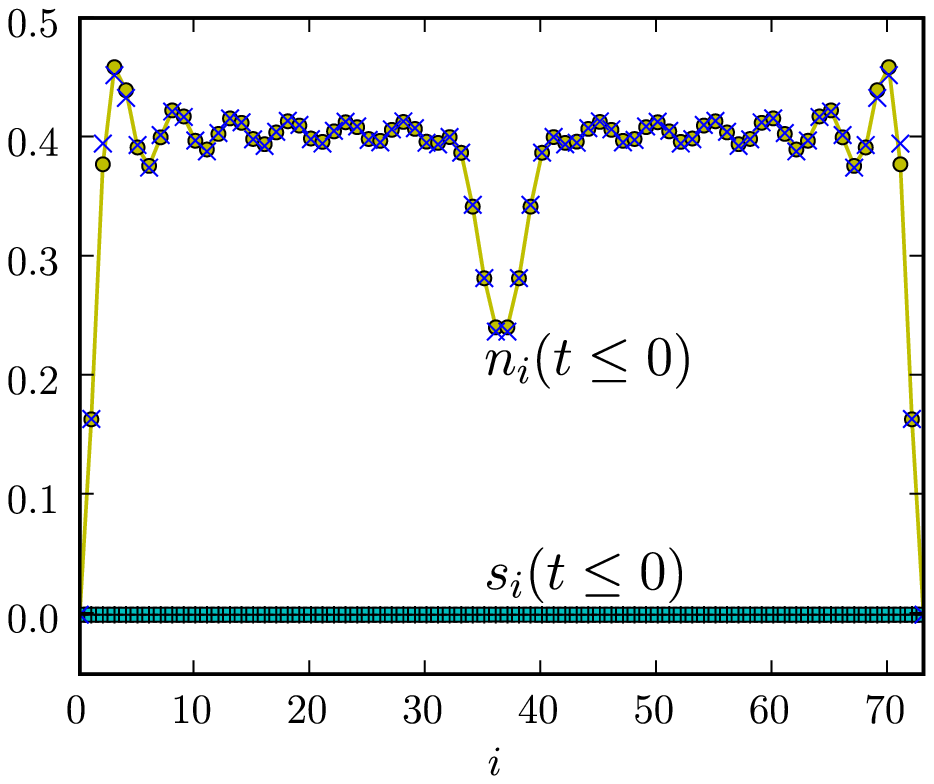}&
\includegraphics[width=0.50\linewidth]{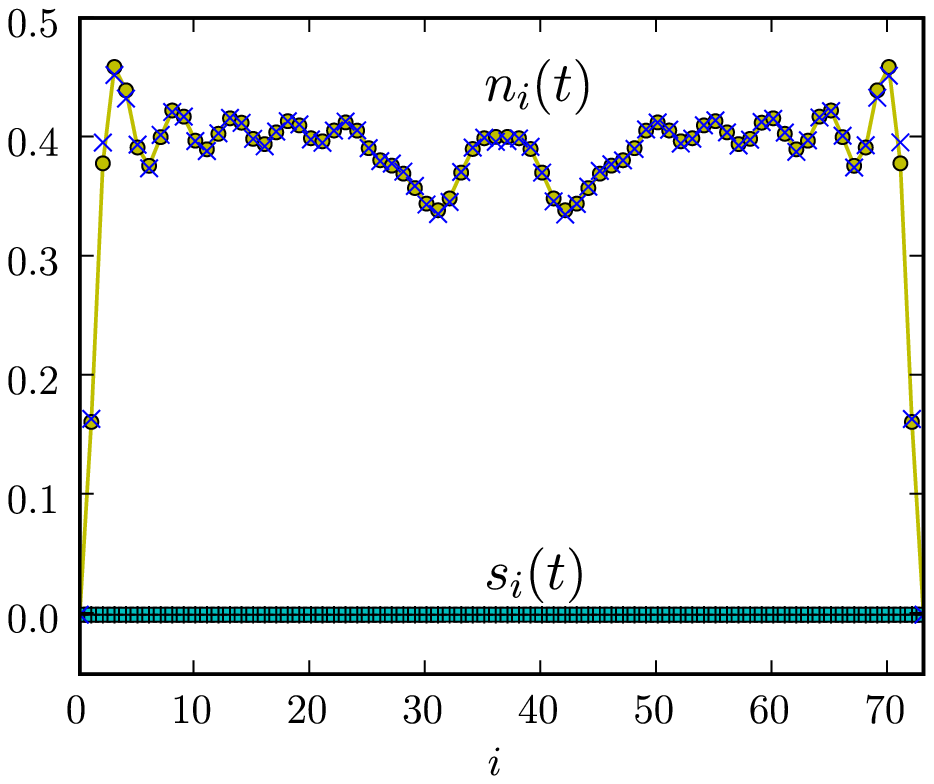}\\
\includegraphics[width=0.50\linewidth]{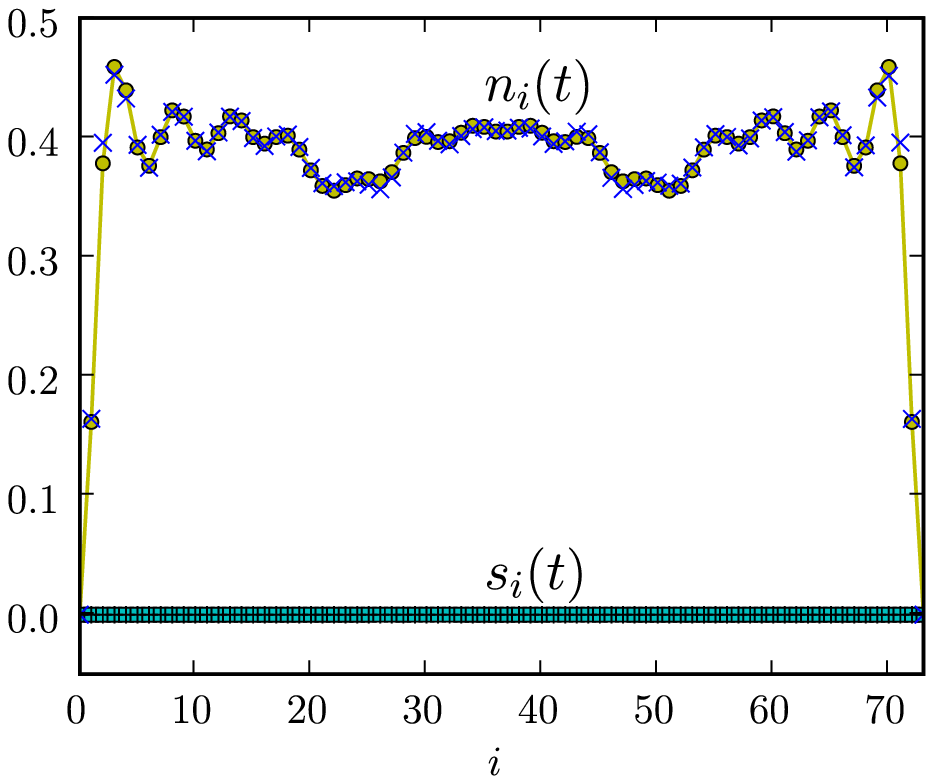}&
\includegraphics[width=0.50\linewidth]{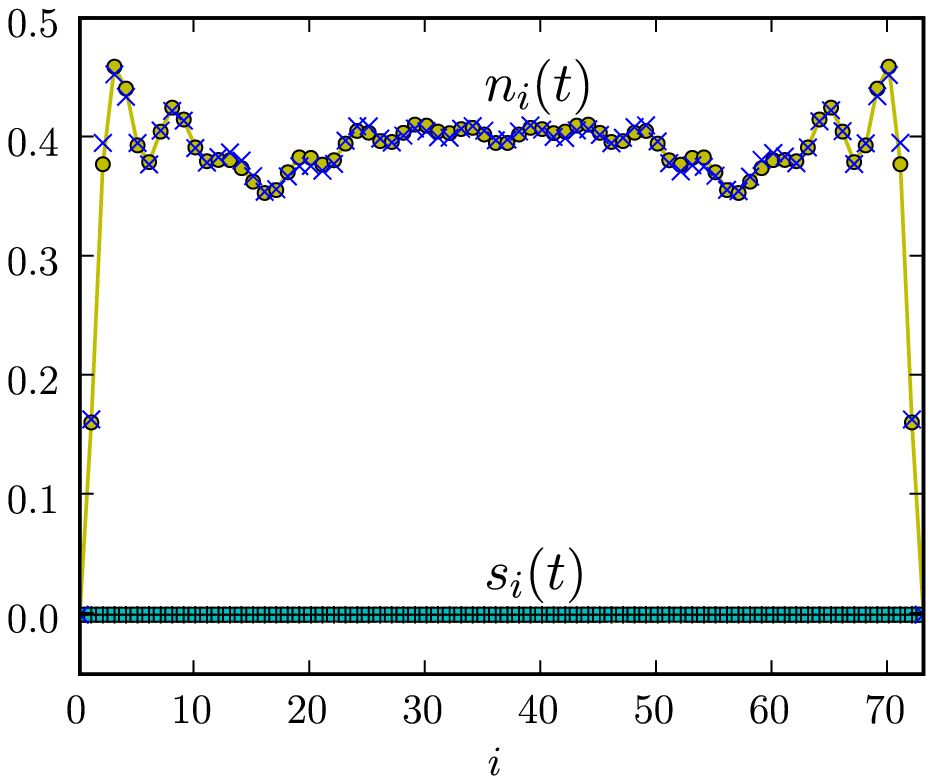}
\end{tabular}
\end{center}
\caption{(Color online) Charge $n_{i}(t)$ and spin $s_i(t)$ occupations as functions of lattice site $i$ 
and time $t$ for $L=72$ (the hard-wall boundary conditions are imposed at the sites $0$ and $73$), $N_\uparrow=N_\downarrow=14$, $u=+2$, $W_\uparrow/\gamma=W_{\downarrow}/\gamma=-7/18$, and $w=2$. TDDFT data (filled circles and squares) are compared with tDMRG data ($\times$ and $+$).
Top left panel: ground-state charge and spin occupations for times $t\leq 0$. Top right panel: same as in the top left panel but at time $t=5~\hbar/\gamma$. Bottom left panel: same as in the top panels but at time $t=10~\hbar/\gamma$. Bottom right panel: same as in the bottom left panel but at time $t=15~\hbar/\gamma$.\label{fig:one}}
\end{figure*}

In Fig.~\ref{fig:one} we show results for a spin-unpolarized system ($N_\uparrow=N_\downarrow=14$) with $U=+2~\gamma$. 
In this special simulation the external potential is chosen to be spin-independent: 
$W_\uparrow=W_\downarrow=-7/18$ [see Eq.~(\ref{eq:ext_pot})]. The initial ``charge" $n_i=n_{i\uparrow }+n_{i\downarrow}$ and ``spin"  local occupations $s_i=n_{i\uparrow}-n_{i\downarrow}$ are shown in the top left panel. The spin occupation is identically zero in this spin-unpolarized situation simply due to the symmetry between the $\uparrow$- and $\downarrow$-spin atoms. The effect of the external potential results in a local perturbation of the charge density distribution, 
{\it i.e.} the dip in Fig.~\ref{fig:one}. 
Additional deviations from a homogeneous distribution are due to OBCs which cause Friedel oscillations. 

\begin{figure*}
\begin{center}
\tabcolsep=0cm
\begin{tabular}{cc}
\includegraphics[width=0.5\linewidth]{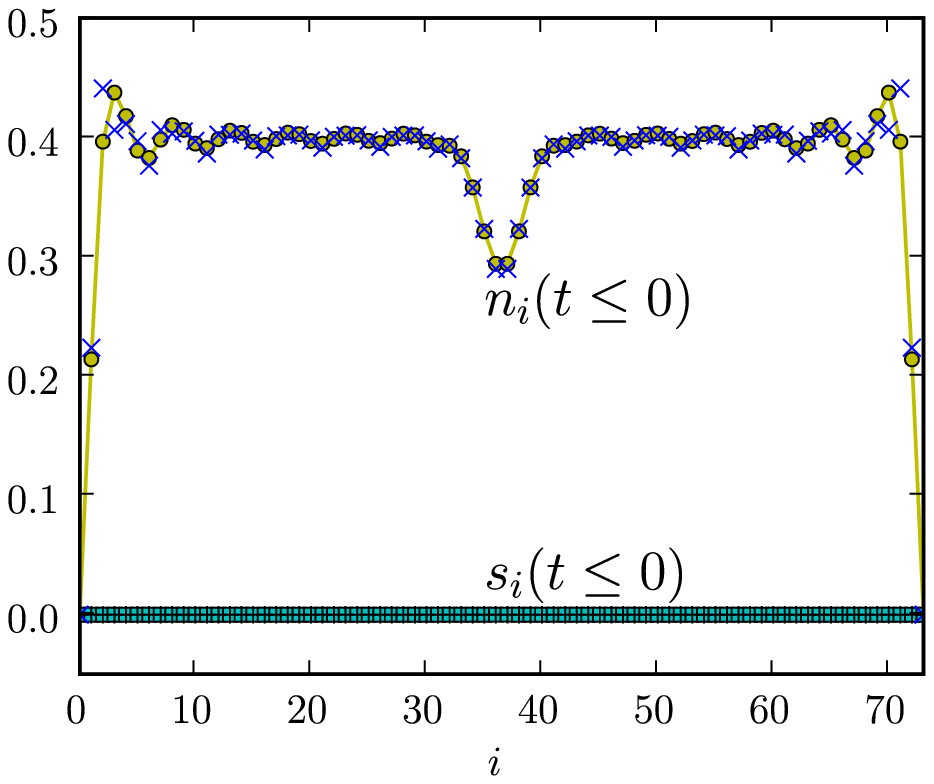}&
\includegraphics[width=0.5\linewidth]{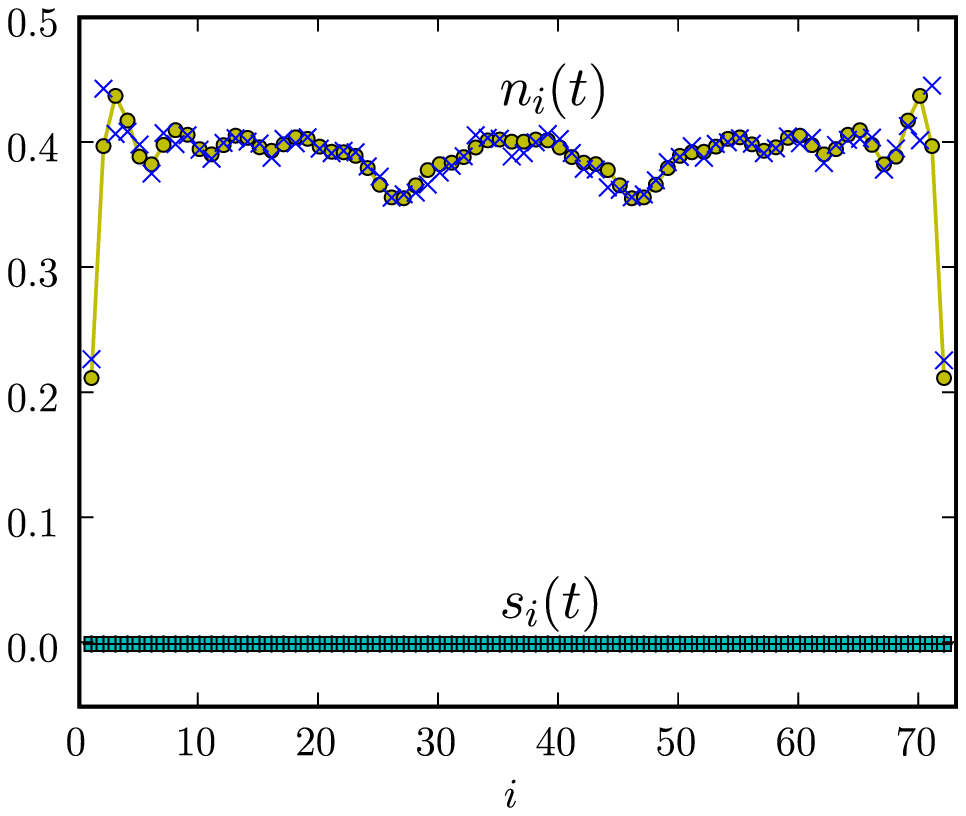}\\
\includegraphics[width=0.5\linewidth]{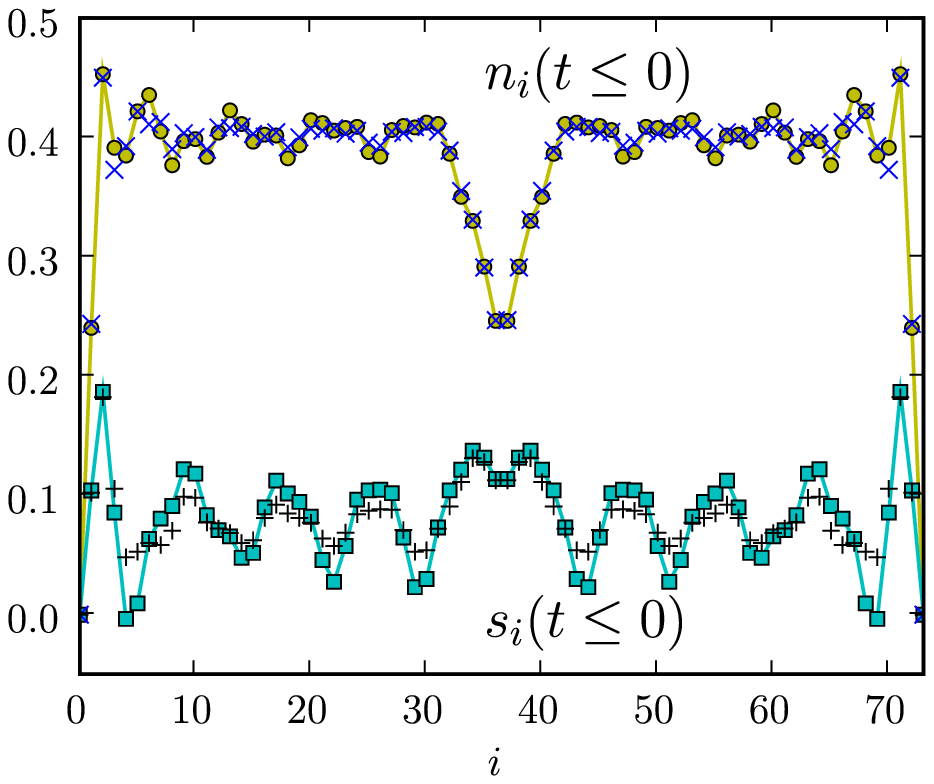}&
\includegraphics[width=0.5\linewidth]{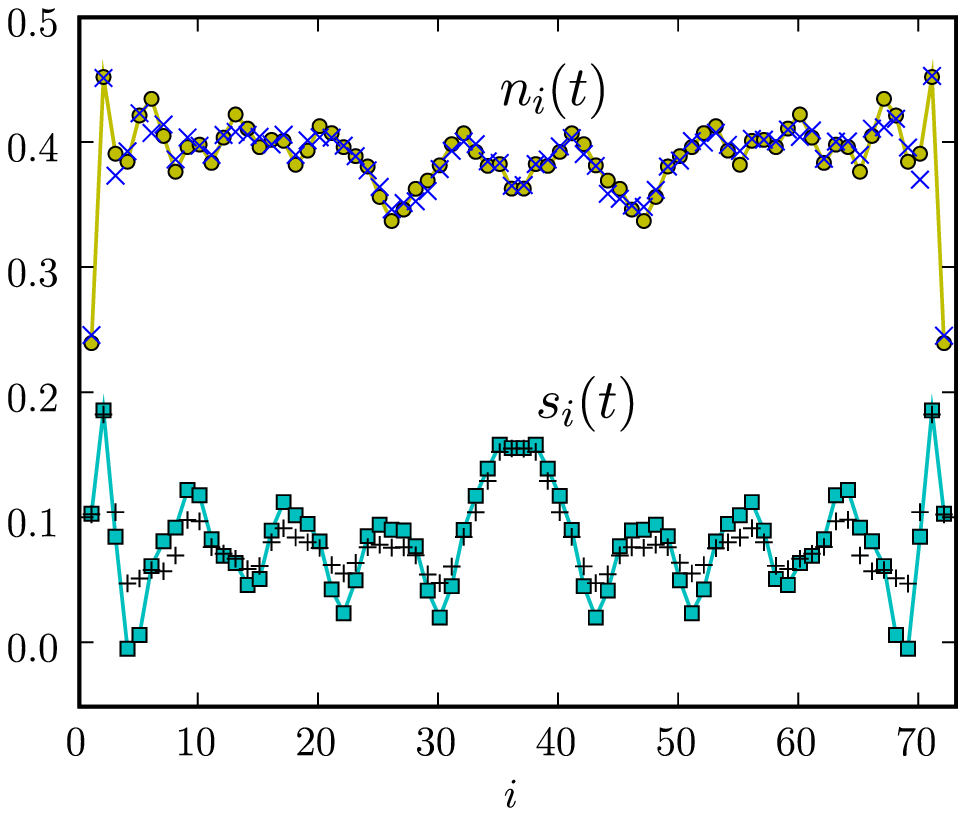}
\end{tabular}
\end{center}
\caption{(Color online) Top panels: 
Charge $n_{i}(t)$ and spin $s_i(t)$ occupations as functions of lattice site $i$ 
and time $t$ for $L=72$, $N_\uparrow=N_\downarrow=14$, $u=+6$, $W_\uparrow/\gamma=W_{\downarrow}/\gamma=-7/18$, and $w=2$. 
TDDFT data (filled circles and squares) are compared with tDMRG data ($\times$ and $+$).
Left: ground-state charge and spin occupations for times $t\leq 0$. 
Right: same as in the left panel but at time $t=6~\hbar/\gamma$. 
Bottom panels: same in the top panels but for $N_\uparrow=20$ and 
$N_\downarrow=8$.\label{fig:six}}
\end{figure*}

We find an excellent agreement between the ground-state DFT/BA-LSDA results (see Sect.~\ref{sect:gs}) 
and the DMRG results. Small differences between the two are visible only close to the boundaries and in the local perturbation. OBCs are very severe boundary conditions for BA-LSDA: the performance of this approximation increases 
in the presence of soft boundaries, such as those created by a parabolic trapping potential---see Ref.~\onlinecite{gao_prb_2006}. 
Our finding is in agreement with earlier studies~\cite{vieira_condmat_2007} 
in which the inability of the BA local-density approximation to reproduce the correct 
Friedel oscillations in the presence of a single impurity (or close to a sharp boundary) has been noted. 
In general, no local-density approximation for the xc potential is expected to produce Friedel oscillations with an amplitude that scales as a power-law as a function of the distance from the impurity/boundary, with an exponent that is controlled by the interaction strength. Of course, this inadequacy of BA-LSDA is more severe at strong coupling: see, for example, the 
top left panel in Fig.~\ref{fig:six}.

The time evolution of charge and spin occupations subsequent to the sudden switching-off 
of the external local potential is illustrated in the other three panels of Fig.~\ref{fig:one} 
for times $t=5, 10$ and $15~\hbar/\gamma$. For small times $t<5~\hbar/\gamma$ the initial dip in the charge density starts splitting into two counter-propagating perturbations. After the splitting the perturbations move with a certain velocity towards the boundaries. During the evolution a deformation of the shape of the density perturbations can be seen. Scattering of the perturbation from the Friedel oscillations and the boundaries occurs. 

\begin{figure}
\begin{center}
\includegraphics[width=1.00\linewidth]{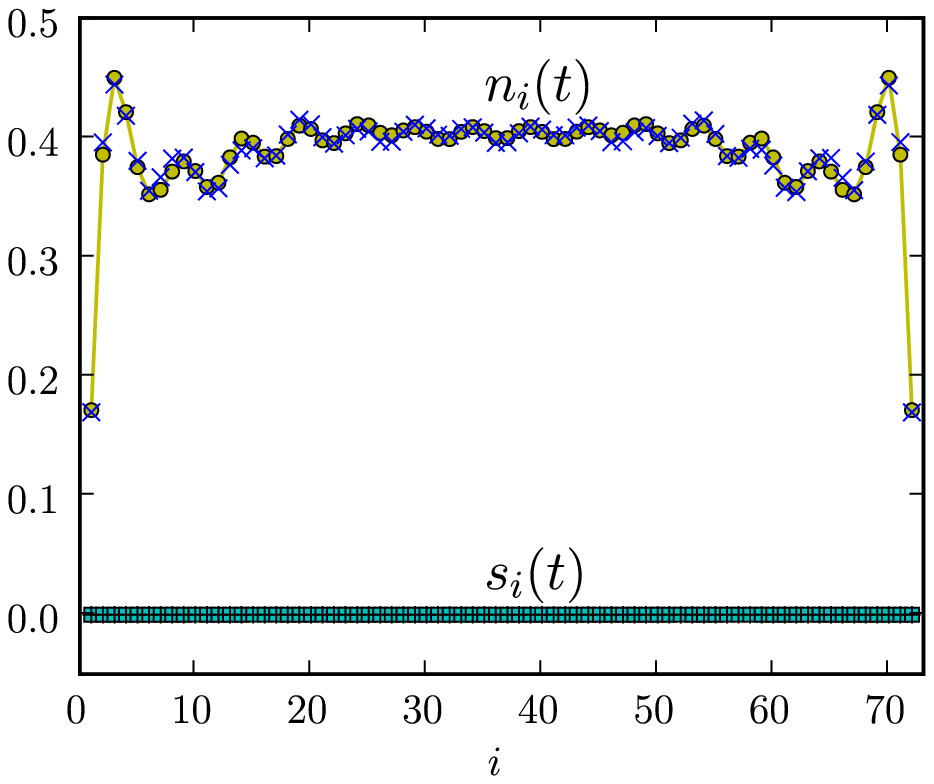}
\includegraphics[width=1.00\linewidth]{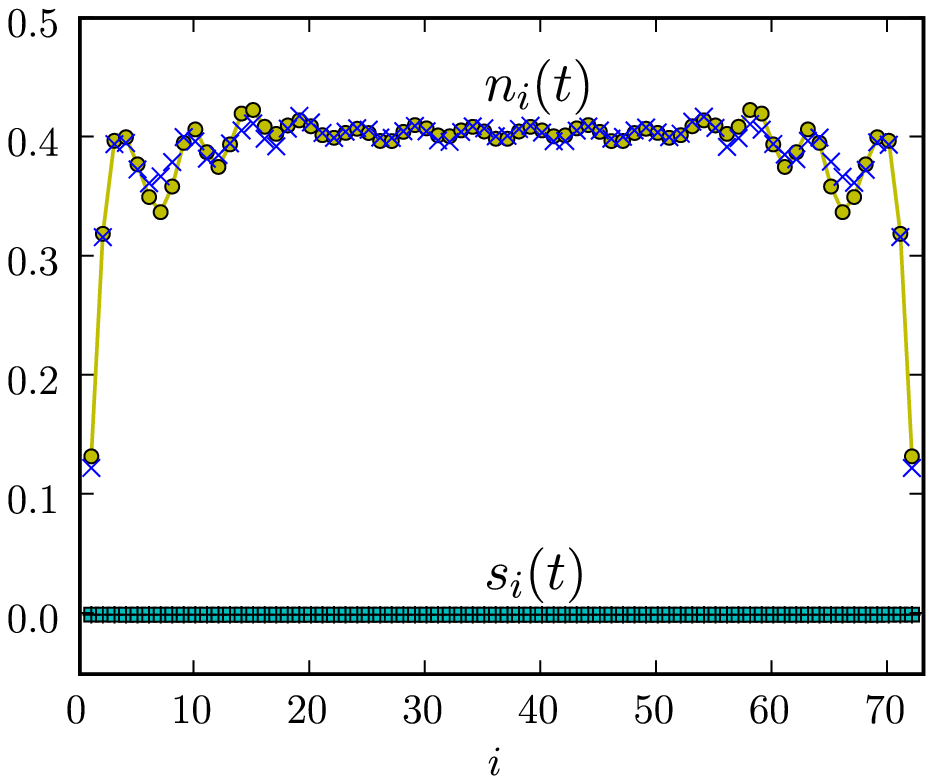}
\end{center}
\caption{(Color online) Same as in Fig.~\ref{fig:one} but for $t=20~\hbar/\gamma$ (top panel) and $t=30~\hbar/\gamma$ (bottom panel).\label{fig:two}}
\end{figure}

We see how TDDFT/BA-ALSDA results are 
in full quantitative agreement with the tDMRG data. The agreement remains quite decent 
even at times later than  $t=15~\hbar/\gamma$, when reflections from the boundaries and 
interference with the microscopic Friedel oscillations are expected to spoil the performance of TDDFT/BA-ALSDA. 
For example, in Fig.~\ref{fig:two} we have reported the comparison between TDDFT/BA-ALSDA and tDMRG data 
at $t=20$ and $30~\hbar/\gamma$.

\begin{figure*}
\begin{center}
\tabcolsep=0cm
\begin{tabular}{cc}
\includegraphics[width=0.50\linewidth]{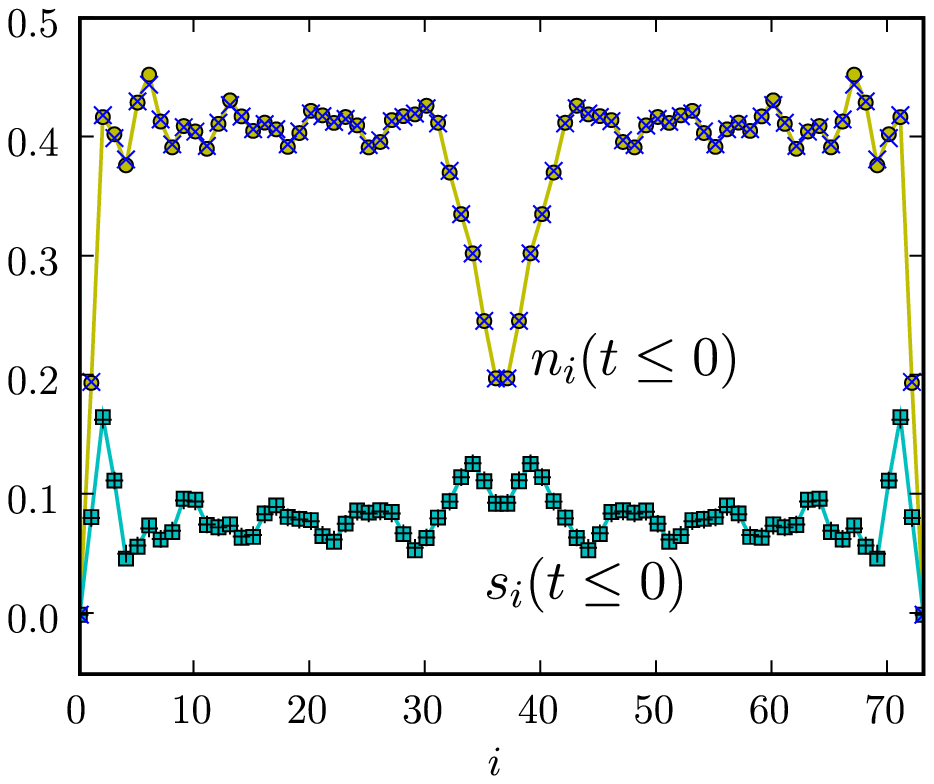}&
\includegraphics[width=0.50\linewidth]{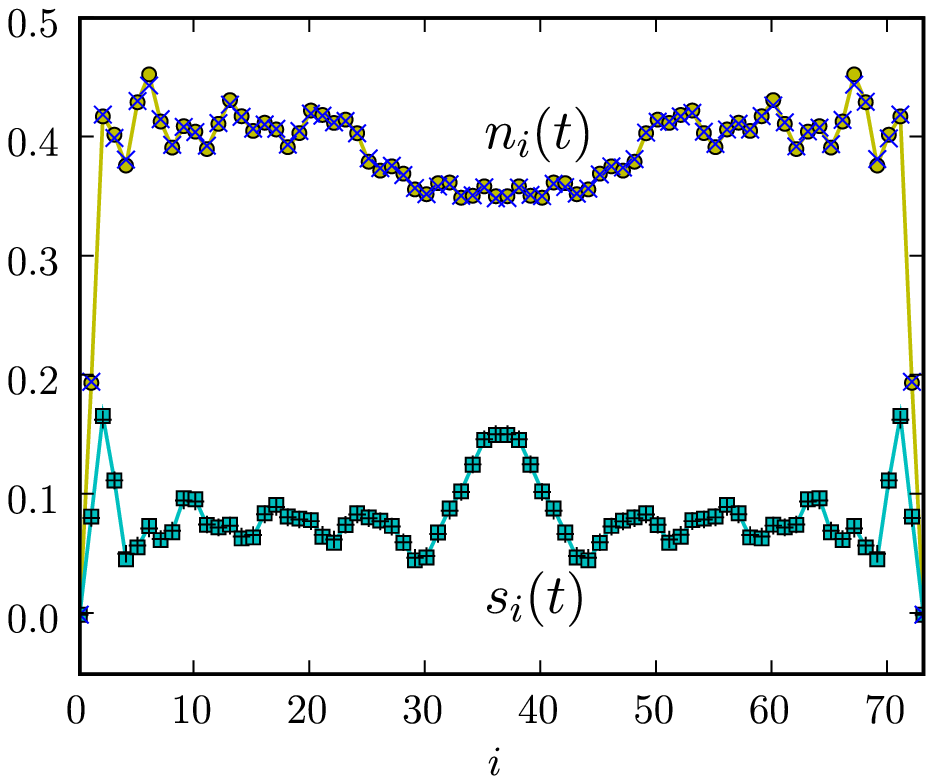}\\
\includegraphics[width=0.50\linewidth]{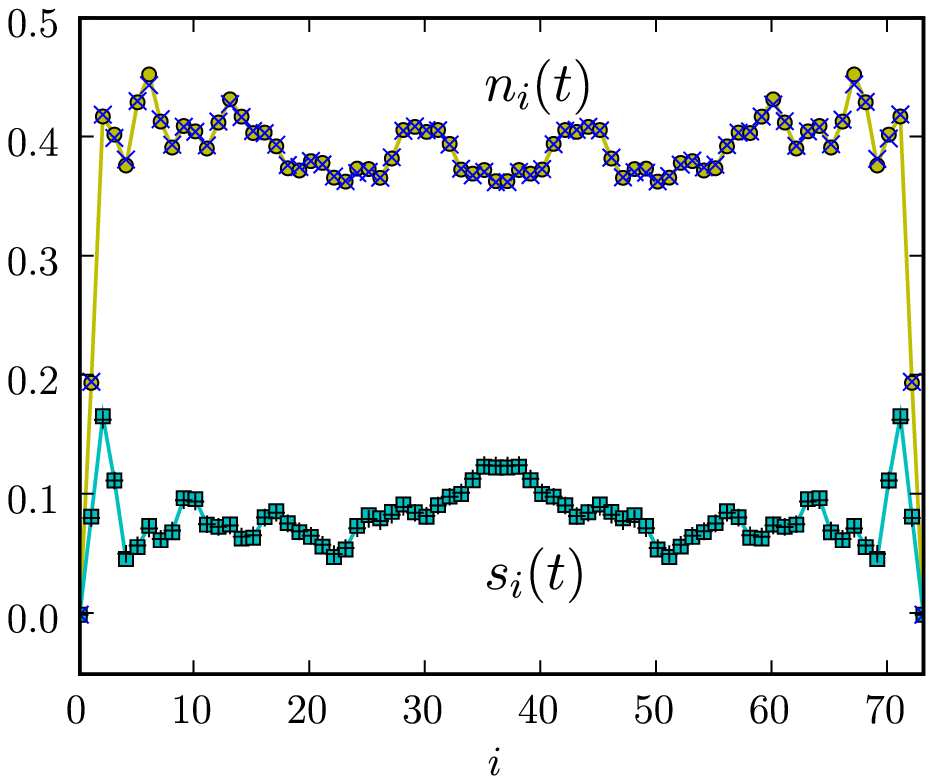}&
\includegraphics[width=0.50\linewidth]{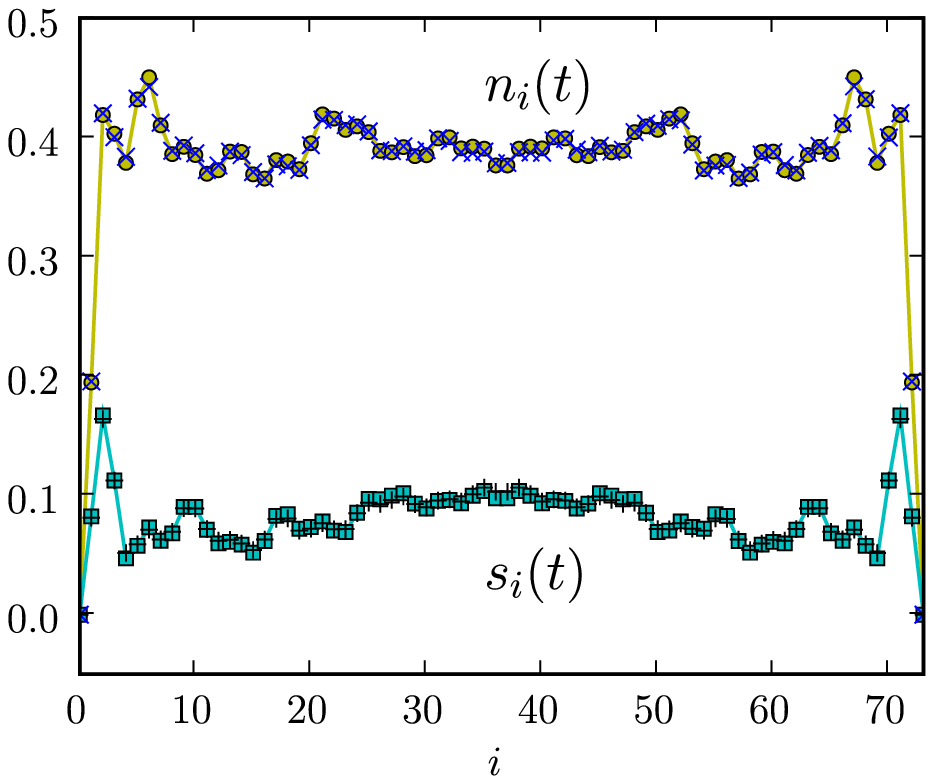}
\end{tabular}
\end{center}
\caption{(Color online) Same as in Fig.~\ref{fig:one} but for 
$N_\uparrow=20$, $N_\downarrow=8$, $u=+2$, $W_\uparrow/\gamma=W_{\downarrow}/\gamma=-5/9$, and $w=2$.\label{fig:three}}
\end{figure*}

In the results shown in Figs.~\ref{fig:one} and~\ref{fig:two} the spin dynamics is trivial: $s_i(t)\equiv 0$ at all times. 
In order to check the predicting power of TDDFT/BA-ALSDA in regard to spin dynamics, in Fig.~\ref{fig:three} we show 
results for a spin-polarized system ($N_\uparrow=20$ and $N_\downarrow=8$). 
Note that the local external potential in this case is again spin-independent (as it was in the case of Fig.~\ref{fig:one}) 
and couples only to the total-charge sector. Nonetheless, due to the imbalance between the number of atoms with different spin, this local disturbance generates a non-trivial spin dynamics. 
This is a consequence of the fact that the spin and charge sectors of 
the spin-polarized 1D Hubbard model away from half-filling are coupled~\cite{Bahder86,Woynarovich,kollath_njp_2006,Frahm07,rizzi_FFLO_2007}. The coupling effects the dynamics considerably. In contrast to the spin-balanced case, the perturbation does not decay into noninteracting charge and spin perturbations. As before the charge-density perturbation evolves into two counterpropagating perturbations. However, the creation of a pronounced dip at the center of the system, which is due to the interaction with the slower spin perturbation, can be observed. 
We clearly see from Fig.~\ref{fig:three} how the agreement between TDDFT/BA-ALSDA and tDMRG is also very satisfactory in this spin-polarized case. Deviations are mainly observed close to the boundaries.

\begin{figure*}
\begin{center}
\tabcolsep=0cm
\begin{tabular}{cc}
\includegraphics[width=0.50\linewidth]{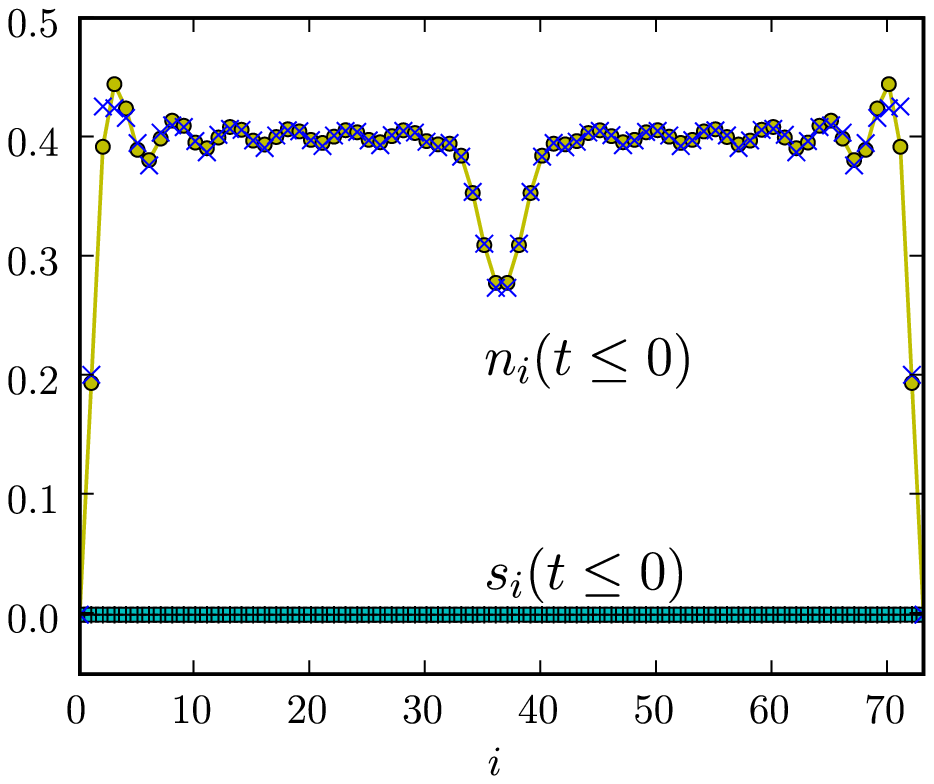}&
\includegraphics[width=0.50\linewidth]{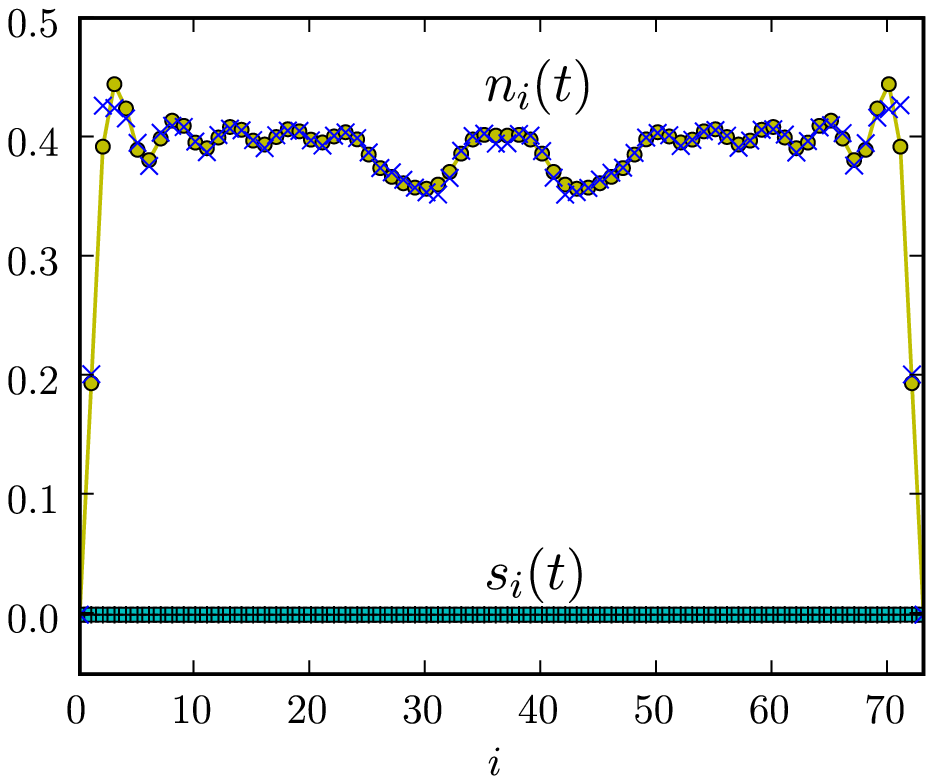}\\
\includegraphics[width=0.50\linewidth]{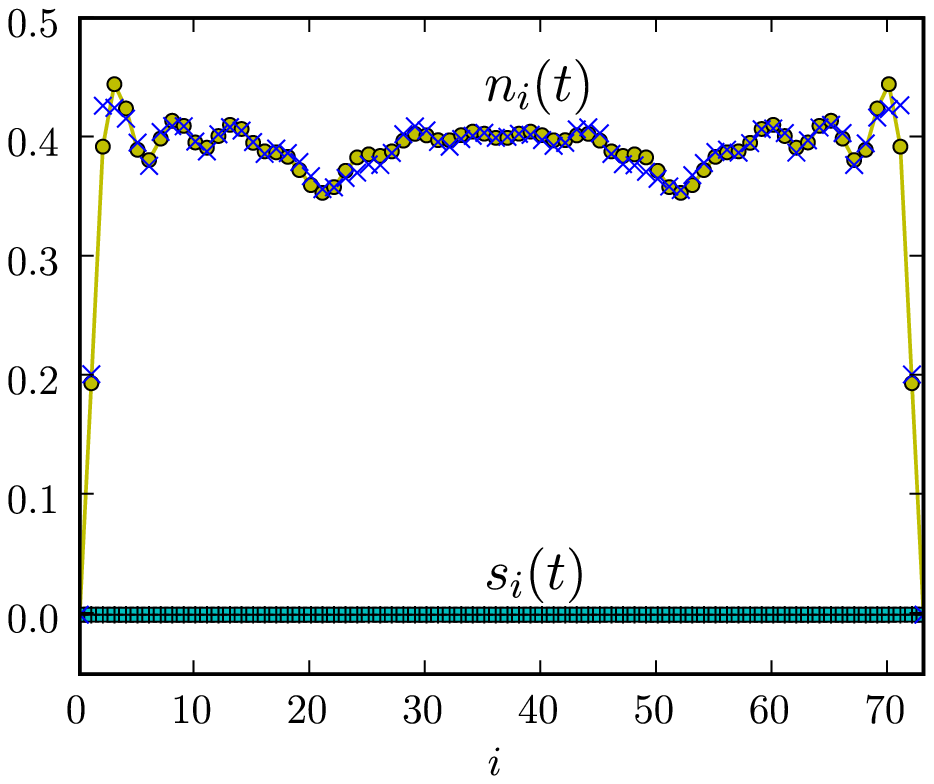}&
\includegraphics[width=0.50\linewidth]{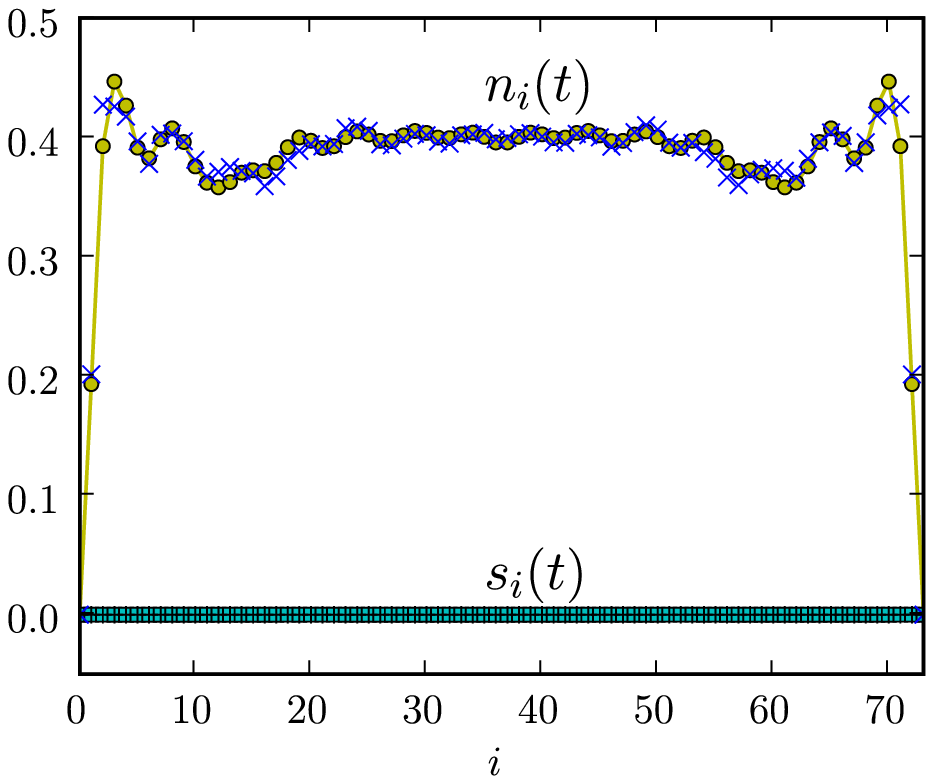}
\end{tabular}
\end{center}
\caption{(Color online) Same as in Fig.~\ref{fig:one} but for $u=+4$.\label{fig:four}}
\end{figure*}

\begin{figure*}
\begin{center}
\tabcolsep=0cm
\begin{tabular}{cc}
\includegraphics[width=0.50\linewidth]{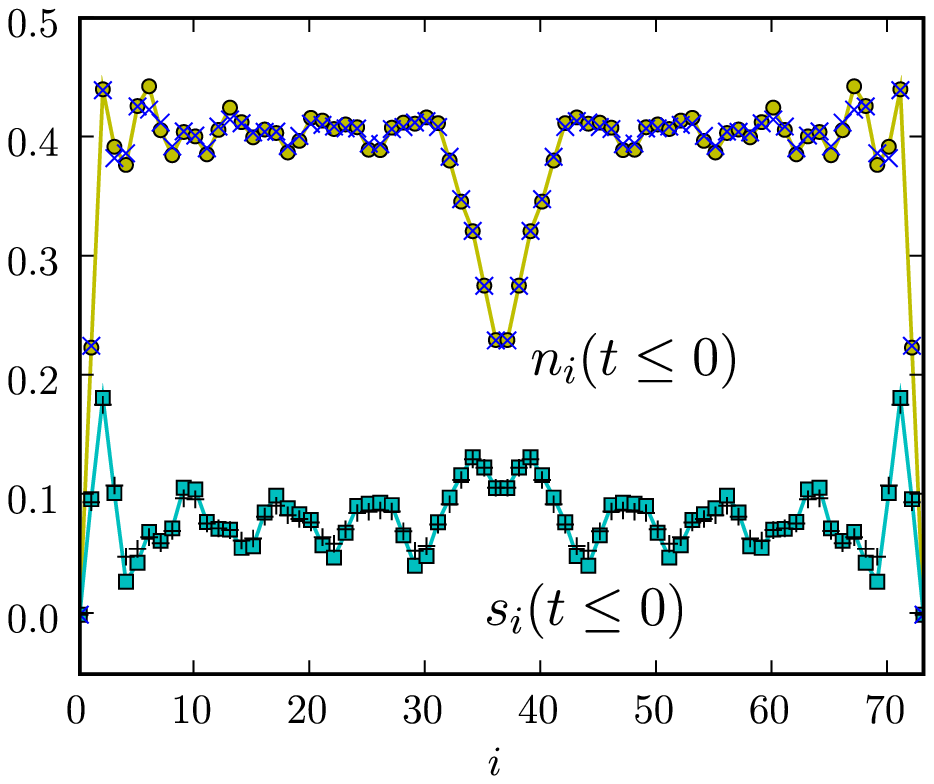}&
\includegraphics[width=0.50\linewidth]{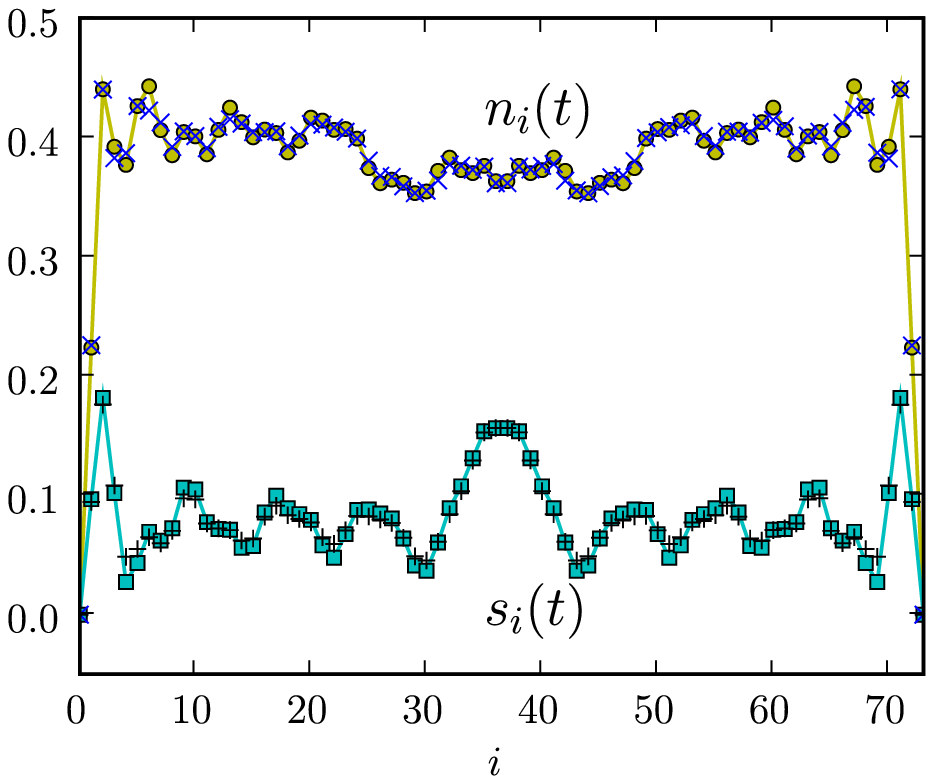}\\
\includegraphics[width=0.50\linewidth]{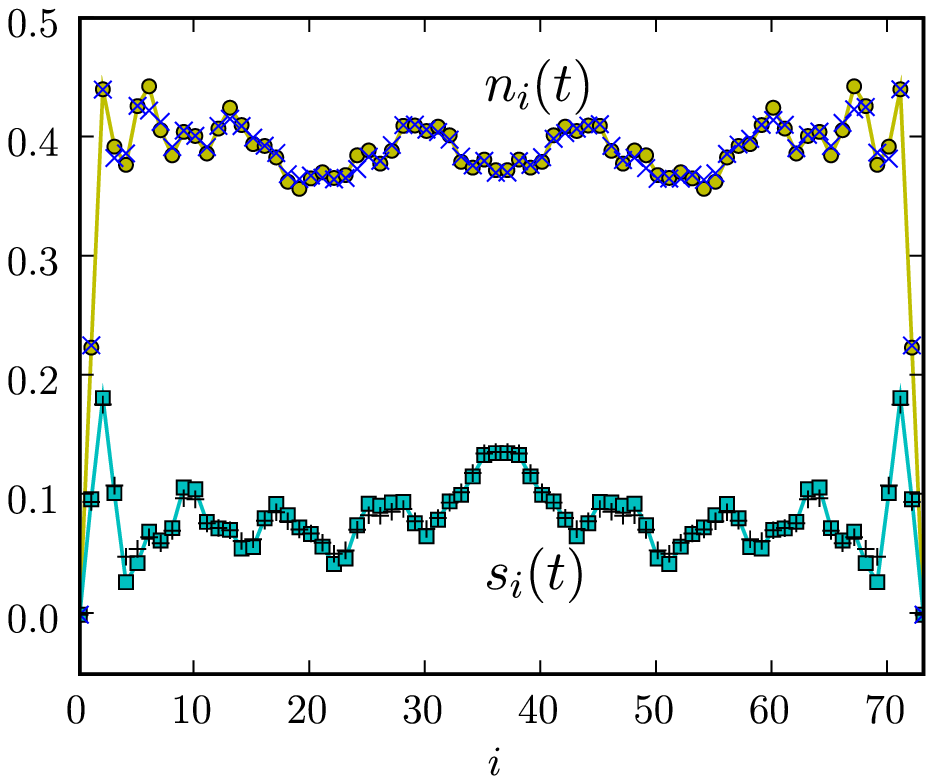}&
\includegraphics[width=0.50\linewidth]{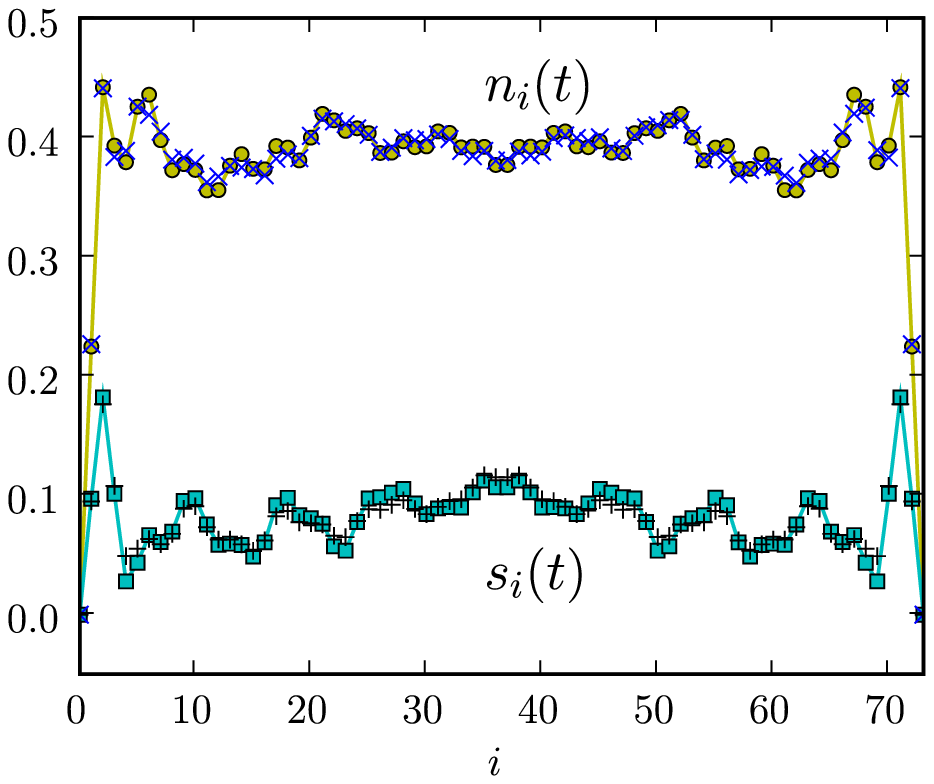}
\end{tabular}
\end{center}
\caption{(Color online) Same as in Fig.~\ref{fig:three} but for $u=+4$.\label{fig:five}}
\end{figure*}

All numerical results shown in Figs.~\ref{fig:one}-\ref{fig:three} have been obtained for $U=+2~\gamma$. 
We have also checked how the relative strength $U/\gamma$ of interactions influences the reliability of TDDFT/BA-ALSDA. 
In Figs.~\ref{fig:four}-\ref{fig:six} we show results for spin-unpolarized and spin-polarized systems 
at $U=+4~\gamma$ and $U=+6~\gamma$. With increasing coupling strength the deviations between the two methods 
increase both for the initial state and the time evolution. For the initial state the Friedel oscillations are not captured correctly (mainly the $4k_{\rm F}$ component~\cite{vieira_condmat_2007}) and large deviations at the boundaries arise. 
In particular, the results for the spin-density profiles at $U=+6~\gamma$ show considerable differences between the two methods. In the time evolution, in contrast to smaller couplings, the form of the charge perturbation shows deviations between the results of the two methods even in the unpolarized case (see the top right panel in Fig.~\ref{fig:six}). 
However, the main features of the tDMRG data are qualitatively reproduced by TDDFT/BA-ALSDA ({\it e.g.} 
the splitting into two counter-propagating perturbations with approximately the correct velocity).
From these plots we can see how the predicting power of TDDFT/BA-ALSDA is quite acceptable also at strong coupling.

\section{Conclusions and future perspectives}
\label{sect:conclusions}

In summary, we have carried out an extensive numerical study of the collective ``spin" and ``charge" dynamics in 
strongly correlated ultracold Fermi gases confined in one-dimensional tight tube. We have compared the results obtained from time-dependent density-functional theory within a suitable (Bethe-{\it Ansatz}-based) adiabatic local-spin-density approximation with accurate results based on the adaptive time-dependent density-matrix renormalization-group method. We have found the simple adiabatic local-spin-density approximation for the time-dependent exchange-correlation potential to be reliable and surprisingly accurate in describing the collective evolution of density and spin wave packets in a wide 
range of coupling strengths and spin polarizations. 

The adiabatic scheme proposed in this work can also be used and tested in many other interesting problems. For example, 
one can study spin-charge dynamics after a local quench in Luther-Emery liquids 
[which can be model by Eq.~(\ref{eq:hubbard}) with $U<0$], Andreev reflection~\cite{daley_prl_2008} 
in ultracold two-component Fermi gases, and, of course, quantum dynamics in the presence of  
truly time-dependent external potentials.

In this last respect, we can anticipate that more sophisticated functionals in which the main ALDA assumptions, 
{\it i.e.} (i) locality in space {\it and} time and (ii) the complete neglect of memory effects, have been relaxed 
may be needed to handle truly time-dependent problems. Recent interesting numerical results~\cite{verdozzi_2007} 
of a TDDFT study of small ({\it i.e.} $L=4$ to $L=12$) Hubbard chains seem indeed to indicate that {\it both} 
(i) and (ii) are ultimately necessary ingredients that the time-dependent xc potential of TDDFT 
should possess for the description of strongly correlated systems.

Finally, we would like to mention that adiabatic and beyond-adiabatic exchange-correlation functionals based on time-dependent current-spin-density-functional theory have also been recently applied~\cite{gao_condmat_2008} to study collective spin and charge dynamics at finite temperature in one-dimensional continuum models.

\acknowledgments
G.X. was supported by NSF of China under Grant No. 10704066. 
C.K. thanks the network ``Triangle de la Physique" and the DARPA OLE program for support. 
G.X. and M.P. acknowledge many useful discussions with Klaus Capelle and Giovanni Vignale.

\begin{thebibliography}{77}

\bibitem{giamarchi_book}
	J. Voit, Rep. Prog. Phys. {\bf 57}, 977 (1994);
	A.O. Gogolin, A.A. Nersesyan, and A.M. Tsvelik, 
	{\it Bosonization and Strongly Correlated Systems} (Cambridge University Press, Cambridge, 1998);
	H.J. Schulz, G. Cuniberti, and P. Pieri, in {\it Field Theories for Low-Dimensional 
	Condensed Matter Systems}, edited by G. Morandi, P. Sodano, A. Tagliacozzo, and V. 
	Tognetti (Springer, Berlin, 2000) p. 9;
	T. Giamarchi, {\it Quantum Physics in One Dimension} (Clarendon Press, Oxford, 2004).

\bibitem{Saito_book} 
	R. Saito, G. Dresselhaus, and M.S. Dresselhaus, {\it Physical Properties of Carbon Nanotubes} 
	(Imperial College Press, London, 1998).

\bibitem{yacoby_sc_separation}
	O.M. Auslaender, A. Yacoby, R. de Picciotto, K.W. Baldwin, L.N. Pfeiffer, and K.W. West, 
	Science {\bf 295}, 825 (2002); 
	O.M. Auslaender, H. Steinberg, A. Yacoby, Y. Tserkovnyak, B.I. Halperin, K.W. Baldwin, L.N. Pfeiffer, and K.W. West, 
	{\it ibid.} {\bf 308}, 88 (2005).

\bibitem{Nitzan_2003} 
  	A. Nitzan and M.A. Ratner, Science {\bf 300}, 1384 (2003).

\bibitem{xLL}
	For a review see {\it e.g.} A.M. Chang, \rmp {\bf 75}, 1449 (2003).
	
\bibitem{rmp_cold_atoms}
	For a recent review see 
	I. Bloch, J. Dalibard, and W. Zwerger, arXiv:0704.3011v2, to be published in \rmp (2008).	
	
\bibitem{moritz_2005}
	H. Moritz, T. St\"oferle, K. G\"unter, M. K\"ohl, and T. Esslinger, \prl {\bf 94}, 210401 (2005). 
	
\bibitem{greiner_2001}
	M. Greiner, I. Bloch, O. Mandel, T.W. H\"ansch, and T. Esslinger, \prl {\bf 87}, 160405 (2001); 
 	H. Moritz, T. St\"oferle, M. K\"ohl, and T. Esslinger, {\it ibid.} {\bf 91}, 250402 (2003);
 	T. St\"oferle, H. Moritz, C. Schori, M. K\"ohl, and T. Esslinger, {\it ibid.} {\bf 92}, 130403 (2004);
 	B.L. Tolra, K.M. O'Hara, J.H. Huckans, W.D. Phillips, S.L. Rolston, and J.V. Porto, {\it ibid.} {\bf 92}, 190401 (2004);
 	B. Paredes, A. Widera, V. Murg, O. Mandel, S. F\"olling, I. Cirac, 
 	G.V. Shlyapnikov, T.W. H\"ansch, and I. Bloch, Nature {\bf 429}, 277 (2004);
 	T. Kinoshita, T. Wenger, and D.S. Weiss, Science {\bf 305}, 1125 (2004).	
	
\bibitem{haldane}
	F.D.M. Haldane, J. Phys. C {\bf 14}, 2585 (1981) and \prl {\bf 47}, 1840 (1981).	

\bibitem{kollath_prl_2005} 
	C. Kollath, U. Schollw\"ock, and W. Zwerger, \prl {\bf 95}, 176401 (2005).

\bibitem{kollath_jpb_2006} 
	C. Kollath, J. Phys. B {\bf 39}, S65 (2006).

\bibitem{kollath_njp_2006}
	C. Kollath and U. Schollw\"ock, New J. Phys. {\bf 8}, 220 (2006).

\bibitem{kecke_prl_2005}
	L. Kecke, H. Grabert, and W. H\"ausler, \prl  {\bf 94}, 176802 (2005).
	
\bibitem{recati_prl_2003}
	A. Recati, P.O. Fedichev, W. Zwerger, and P. Zoller, \prl {\bf 90}, 020401 (2003) and J. Opt. B {\bf 5}, S55 (2003).

\bibitem{polini_PRL_2007}
	M. Polini and G. Vignale, \prl {\bf 98}, 266403 (2007);
	D. Rainis, M. Polini, M.P. Tosi, and G. Vignale, \prb {\bf 77}, 035113 (2008).

\bibitem{feiguin}
	S.R. White and A.E. Feiguin, \prl {\bf 93}, 076401 (2004).
	
\bibitem{daley}
	A.J. Daley, C. Kollath, U. Schollw\"ock, and G. Vidal, J. Stat. Mech. P04005 (2004).
	
\bibitem{Giuliani_and_Vignale}
	G.F. Giuliani and G. Vignale, {\it Quantum Theory of the Electron Liquid} 
	(Cambridge University Press, Cambridge, 2005).	

\bibitem {vignale_kohn}
	G. Vignale and W. Kohn, in {\em Electronic Density Functional Theory}, 
	edited by J. Dobson, M. K. Das, and G. Vignale (Plenum Press, New York, 1996).
	
\bibitem {marques_2006}
	{\it Time-Dependent Density Functional Theory}, Lecture Notes in Physics Vol. {\bf 706}, 
	edited by M.A.L. Marques, F. Noguiera, A. Rubio, K. Burke, C.A. Ullrich, and E.K.U. Gross (Springer, Berlin, 2006).
	
\bibitem{rgt} 
	E. Runge and E.K.U. Gross, \prl {\bf 52}, 997 (1984); see also R. van Leeuwen, 
	{\it ibid.} {\bf 82}, 3863 (1999).

\bibitem {zangwill}  
	A. Zangwill and  P. Soven,  \prl {\bf 45}, 204 (1980); \prb {\bf 24}, 4121 (1981).

\bibitem{vk_1996}
	G. Vignale and W. Kohn, \prl {\bf 77}, 2037 (1996).
	
\bibitem{dobson_1997}
	J.F. Dobson, M.J. B\"{u}nner, and E.K.U. Gross, \prl {\bf 79}, 1905 (1997).
	
\bibitem{vuc_1997}
	G. Vignale, C.A. Ullrich, and S. Conti, \prl {\bf 79}, 4878 (1997).
	
\bibitem{tokatly_2005}
	I.V. Tokatly, \prb {\bf 71}, 165104 and 165105 (2005); {\it ibid.} {\bf 75}, 125105 (2007).
	
\bibitem{orestes_2007}	
	E. Orestes, K. Capelle, A.B.F. da Silva, and C.A. Ullrich, J. Chem. Phys. {\bf 127}, 124101 (2007).		
	
\bibitem{soft}
	O. Gunnarsson and K. Sch\"onhammer, \prl {\bf 56}, 1968 (1986);
	K. Sch\"onhammer and O. Gunnarsson, J. Phys. C {\bf 20}, 3675 (1987) and \prb {\bf 37}, 3128 (1988); 
	K. Sch\"onhammer, O. Gunnarsson, and R.M. Noack, \prb {\bf 52}, 2504 (1995).	

\bibitem{capelle}
	N.A. Lima, L.N. Oliveira, and K. Capelle, Europhys. Lett. {\bf 60}, 601 (2002);
	N.A. Lima, M.F. Silva, L.N. Oliveira, and K. Capelle, \prl {\bf 90}, 146402 (2003); 
	K. Capelle, N.A. Lima, M.F. Silva, and L.N. Oliveira, in 
	{\it The fundamentals of electron density, density matrix and density functional theory in atoms, 
	molecules and solids}, edited by N.I. Gidopoulos and S. Wilson,  
	Kluwer series ``Progress in Theoretical Chemistry and Physics'' (Kluwer, Dordrecht, 2003);
	P.E.G. Assis, V.L. L\'\i bero, and K. Capelle, \prb {\bf 71}, 052402 (2005);
	M.F. Silva, N.A. Lima, A.L. Malvezzi, and K. Capelle, {\it ibid.} {\bf 71}, 125130 (2005).

\bibitem{gao_prb_2006}
	Gao Xianlong, M. Polini, M.P. Tosi, V.L. Campo, K. Capelle, and M. Rigol, \prb {\bf 73}, 165120 (2006);
	see also Gao Xianlong, M. Polini, B. Tanatar, and M. P. Tosi, {\it ibid.} {\bf 73}, 161103 (2006);
	Gao Xianlong, M. Polini, R. Asgari, and M. P. Tosi, \pra {\bf 73}, 033609 (2006);
	Gao Xianlong, M. Rizzi, M. Polini, R. Fazio, M.P. Tosi, V.L. Campo, Jr., and K. Capelle, 
	\prl {\bf 98}, 030404 (2007).
	
\bibitem{schenk_condmat_2008}
	S. Schenk, M. Dzierzawa, P. Schwab, and U. Eckern, arXiv:0802.2490v1.

\bibitem{verdozzi_2007}
	C. Verdozzi, arXiv:0707.2317v1.		
	
\bibitem{lieb_wu}
	E.H. Lieb and F.Y. Wu, \prl {\bf 20}, 1445 (1968); see also H. Shiba, \prb {\bf 6}, 930 (1972).
	
\bibitem{jaksch_1998}
	D. Jaksch, C. Bruder, J.I. Cirac, C.W. Gardiner, and P. Zoller, \prl {\bf 81}, 3108 (1998);
	W. Hofstetter, J.I. Cirac, P. Zoller, E. Demler, and M.D. Lukin, \prl {\bf 89}, 220407 (2002).	
	

\bibitem{d&g}
	R.M. Dreizler and E.K.U. Gross, {\it Density Functional Theory} (Springer, Berlin, 1990).

\bibitem{joulbert_1998}
	{\it Density Functionals: Theory and Applications}, edited by D. Joulbert, Springer Lecture Notes in Physics 
	Vol.~ {\bf 500} (Springer, Berlin, 1998).
	
\bibitem{vieira_condmat_2007}
	D. Vieira, H.J. P. Freire, V.L. Campo, Jr., and K. Capelle, arXiv:0710.0358v1 
	(to appear in J. Magn. Magn. Mater.).	
	
\bibitem{Gobert}
	D. Gobert, C. Kollath, U. Schollw\"ock, and G. Sch\"utz, Phys. Rev. E {\bf 71}, 036102 (2005).

\bibitem{Bahder86}
	T.B. Bahder and F. Woynarovich, \prb {\bf 33}, 2114 (1986).

\bibitem{Woynarovich}
	F. Woynarovich, \prb {\bf 43}, 11448 (1991); 
	K. Penc and F. Woynarovich, Z. Phys. {\bf 85}, 269 (1991).
	
\bibitem{Frahm07}
	H. Frahm and T. Vekua, J. Stat. Mech. P01007 (2008).
	
\bibitem{rizzi_FFLO_2007}
	M. Rizzi, M. Polini, M.A. Cazalilla, M.R. Bakhtiari, M.P. Tosi, and R. Fazio, arXiv:0712.3364v1 
	(to appear in Phys. Rev. B).
	
\bibitem{daley_prl_2008}
	A.J. Daley, P. Zoller, and B. Trauzettel, \prl  {\bf 100}, 110404 (2008).	
			
\bibitem{gao_condmat_2008}
	Gao Xianlong, M. Polini, D. Rainis, M.P. Tosi, and G. Vignale, arXiv:0804.1514v1. 	
		
\end{thebibliography}
\end{document}